\documentstyle[12pt]{article}

\newcommand{\fig}[1]{Fig.\ref{#1}}
\newcommand{\bq}{\begin{equation}}
\newcommand{\eq}{\end{equation}}
\newcommand{\bqa}{\begin{eqnarray}}
\newcommand{\eqa}{\end{eqnarray}}
\newcommand{\ben}{\begin{enumerate}}
\newcommand{\een}{\end{enumerate}}
\newcommand{\eqn}[1]{Eq.(\ref{#1})}

\def\a2n{${\cal A}(2\nolinebreak \to n)$}
\def\an{${\cal A}(1\to n)$}

\begin{document}
\pagestyle{empty}
\begin{flushright}CERN-TH.6762/92\\
\end{flushright}
\vspace*{1cm}
\begin{center}\begin{Large}
{\bf On amplitude zeros at threshold}\\

\vspace*{2cm}

E.N.~Argyres$^*$ and
Costas~G.~Papadopoulos\\
TH Division, CERN, Geneva, Switzerland\\
\vspace{\baselineskip}
Ronald~H.P.~Kleiss,\\
NIKHEF-H, Amsterdam, the Netherlands\\
\end{Large}
\vspace*{3cm}
Abstract\end{center}
The occurrence
of zeros of \a2n amplitudes at threshold
in scalar theories is studied.
We find a differential equation for the scalar potential,
which incorporates all known cases
where the \a2n amplitudes at threshold vanish for
all sufficiently large $n$,
in all space-time dimensions, $d\ge 1$.
This equation is related to the reflectionless potentials of
Quantum Mechanics and to integrable theories in 1+1 dimensions.
As an application, we find that the sine-Gordon potential and
its hyperbolic version, the sinh-Gordon potential,
also have amplitude zeros at threshold, ${\cal A}(2\to n)=0$,
for $n\ge 4$ and $d\ge 2$,
independently of the mass and the coupling constant.
\vspace{1cm}
\begin{flushleft} CERN-TH.6762/92\\December 1992\\
\vfill
\begin{small}
$*$ On leave of absence from the Institute of Nuclear Physics,
NRCPS
`$\Delta  \eta  \mu  \acute{o}  \kappa  \varrho
  \iota  \tau  o  \varsigma$',
GR-153 10 Athens, Greece.
\end{small}
\end{flushleft}

\newpage
\pagestyle{plain}
\setcounter{page}{1}

The behaviour of amplitudes for processes with a large
number of particles in the final state, originally connected
to the problem of baryon-number violating processes at very high
energies, has recently been studied in the context of
normal perturbation theory as well \cite{gold,vol1,akp1}.
Several properties of scalar
tree-order amplitudes have been established:
\ben
\item
The factorial growth of the
amplitude of one virtual Higgs to go to $n$ on-shell Higsses,
\an, at threshold has been
found to be a generic property of monomial scalar theories, like
$\frac{\lambda}{m!}\phi^m$, for $m\ge 3$ \cite{vol1,akp1},
as well as of the spontaneously broken $\phi^4$ theory.
\item A lower bound on the tree-order cross section
$\sigma(f\bar f\to H^{\ast} \to nH)$ can be established
rigorously, yielding a unitarity-violating behaviour
\cite{vol2,akp2}.
\een
The construction of a theory in which these amplitudes do not
grow factorially with $n$ has been performed in
\cite{akp3}. In such a theory the potential
differs from the free-field one only by factors of
$\log\phi$, indicating that the unitarity-violation
problem emerges when the anharmonic terms of the potential
dominate over the harmonic one (as they will in any finite polynomial
potential).
The factorial growth is related to the
radius of convergence of the generating function of the amplitudes
\cite{akp3},
or equivalently to the pole structure of the classical
space-independent field configurations \cite{vol5} in the
complex-time plane.

\par Trying to go beyond tree order, Voloshin \cite{vol3, vol4}
discovered
the phenomenon of {\em nullification\/}: for all $n$ larger than some
$n_0$, the \a2n amplitudes vanish when the final-state scalars are
produced at rest.
In $\phi^4$ theory, $n_0$ was found to be 4.
The explicit form of the amplitudes \a2n
for any $\phi^m$ scalar theory was given in ref.\cite{akp4}, where the
nullification of \a2n amplitudes
was found also for the $\phi^3$, with $n_0=3$, and for the
broken symmetry $\phi^4$ theory, with $n_0=2$ (see also ref.\cite{smit}).
This property
is independent of the self-coupling or the mass of the
scalar particle or the number of space-time dimensions
(provided that $d\ge 2$); it
is related only to the form of the potential.
Nevertheless, it is not a trivial effect, as for instance the
nullification of all ${\cal A}(2\to 2k+1)$ tree-order amplitudes
in a $\phi^{2n}$ theory, $n,k$ being integers, which holds for
all kinematical configurations.
Amplitude zeros emerge also in the cases when the $\phi$ field is
coupled to other boson and/or fermion fields \cite{vol4,brow1}.
As we will see below, this nullification requires definite relations
between the masses and the couplings appearing in the Lagrangian.

In all known cases, nullification occurs
when the form of the second derivative of the potential
evaluated at the classical background corresponds to the
well-known reflectionless potentials of Quantum Mechanics \cite{refl}.
In this note, we establish a differential equation involving only
the potential and its derivatives with respect to the field,
which, if satisfied, leads to nullification.

Let us assume a general scalar-field potential
whose expansion (apart from the mass term and setting $m_H=1$)
is given by
\bq
V(\phi)=\sum\limits_{m=3}^{\infty} {\lambda_m\over m!} \phi^m\;.
\eq
The $a_2(n)\equiv
{\cal A}(H(p_1)+H(p_2) \to nH(q))$
amplitudes are given by the following recursion formula (see \fig{one}):

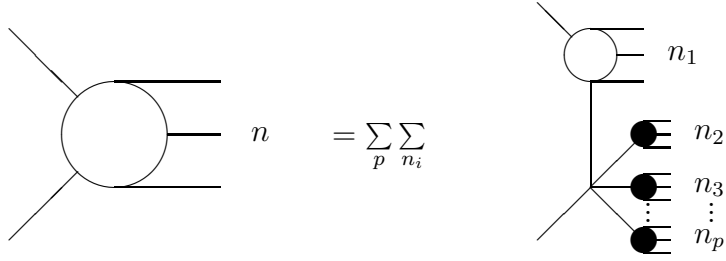
\begin{figure}
\unitlength=2pt
\begin{center}
\begin{picture}(150,60)
\put(50,20){\line(-1,0){20}}
\put(50,30){\line(-1,0){10}}
\put(50,40){\line(-1,0){20}}

\put(10,50){\line(1,-1){13}}
\put(10,10){\line(1,1){13}}
\put(30,30){\circle{20}}

\put(55,25){\makebox(5,10)[c]{$n$}}

\put(70,20){\makebox(20,15)[c]{$ =\sum\limits_{p}\sum\limits_{n_i}$}}

 \put(130,40){\line(-1,0){10}}
 \put(130,45){\line(-1,0){5}}
 \put(130,50){\line(-1,0){10}}
\put(110,55){\line(1,-1){6.5}}
\put(120,45){\circle{10}}
\put(135,40){\makebox(5,10)[c]{$n_1$}}

\put(110,10){\line(1,1){10}}
\put(120,20){\line(0,1){20}}
\put(120,20){\line(1,1){10}}
\put(120,20){\line(1,0){10}}
\put(120,20){\line(1,-1){10}}

\put(135,27.5){\line(-1,0){5}}
\put(135,30){\line(-1,0){2.5}}
\put(135,32.5){\line(-1,0){5}}
\put(130,30){\circle*{5}}
\put(140,27.5){\makebox(5,5)[c]{$n_2$}}

\put(135,22.5){\line(-1,0){5}}
\put(135,20){\line(-1,0){2.5}}
\put(135,17.5){\line(-1,0){5}}
\put(130,20){\circle*{5}}
\put(140,17.5){\makebox(5,5)[c]{$n_3$}}

\put(135,12.5){\line(-1,0){5}}
\put(135,10){\line(-1,0){2.5}}
\put(135,7.5){\line(-1,0){5}}
\put(130,10){\circle*{5}}
\put(140,7.5){\makebox(5,5)[c]{$n_p$}}

\put(130,13.5){.}
\put(130,15){.}
\put(130,16.5){.}

\put(142,13.5){.}
\put(142,15){.}
\put(142,16.5){.}
\end{picture}
\end{center}
\caption[.]{Diagrammatic representation of the recursion formula of
\protect{\eqn{recursa}}. The black circles correspond to the \an
amplitudes at threshold.}
\label{one}
\end{figure}

\bq
a_2(n)
 =  -i\sum\limits_{p=2}^{n}{\lambda_{p+1}\over(p-1)!}
\sum\limits_{\begin{scriptsize}
\begin{array}{c}
n_{2},\ldots,n_{p}\ge 1 \\  n_{1}+\ldots +n_{p}=n
\end{array}
\end{scriptsize} }
{ia_2(n_1)\over P(n_1)}
{ia(n_2)\over (n_2^2-1)}\cdots
{ia(n_p)\over (n_p^2-1)}{n!\over n_1!n_2!\ldots n_p!}\;\;,
\label{recursa}
\eq
\noindent
where $q^\mu=(1,\vec 0)$, $p_1^\mu=(E;\vec p), p_1^2=1$
(recall that $m_H=1$),
$a(n)\equiv {\cal A}(1\to n)$ is
the amplitude at threshold
and $P(n)$ is the inverse propagator, given by
\bq
P(n) = \left(p_1-nq\right)^2-1 = n(n-2 E)\;\;.
\eq
The ans\"{a}tze
\bq
a(n) = -in!(n^2-1)b(n)\;\;\;,\;\;\;
a_2(n) = -in!P(n)b_2(n)\;\;,
\eq
and the introduction of the generating functions
\bq
f(x)=\sum\limits_{n\ge1}b(n)x^n\;\;\;,\;\;\;
f_2(x)=\sum\limits_{n\ge0}b_2(n)x^{n+1}\;\;  \;
\eq
transform \eqn{recursa} into the following
differential equation for $f_2(x)$:
\bq
x^2f_2''(x)-(1+2E) xf_2'(x)-\left[-1-2E+ V''(f(x))\right]f_2(x)
=0\;\;,
\label{difeq}
\eq
with initial conditions $f_2(0)=0$ and $f'_2(0)=1$.
The corresponding equation for $f(x)$ is  \cite{akp1}
\bq
x^2f''(x)+xf'(x)=f(x)+V'(f(x))\;\;,
\label{gen0}
\eq
with initial conditions $f(0)=0$ and $f'(0)=1$.
Equation (\ref{gen0})
can be viewed as the classical equation \cite{brow2} for the
field $\phi(\tau)=f(x)$ with the assignment $x=ce^{\tau}$, $\tau$
being the imaginary time. It can also be interpreted as the static
soliton-like configuration in $1+1$ dimensions, $(t,z)$,
where $x=ce^{z}$ (the constant $c$ guarantees the appropriate
initial conditions for the $f(x)$, at $x=0$, $z=-\infty$).

Let us now assume that $V''(f(x))$ can be written as
\bq
V''=-{ R\over cosh^2(A\tau) }\;\;,
\label{refp}
\eq
where
\bq
\tau=\int_{\phi(0)}^{\phi(\tau)} {d\phi \over
\sqrt{\phi^2+2V(\phi)} }\;.
\label{tau}
\eq
Then, \eqn{difeq} takes the form
\bq
\biggr({d^2\over d\tau^2}-E^2+{ R \over cosh^2(A\tau) }
\biggl)\psi(\tau)=0
\label{ceneq}
\eq
where $x=ce^\tau$ and $f_2(x)=e^{(E+1)\tau}\psi(\tau)$.

This is just
the quantum-mechanical problem of a reflectionless potential:
we must have an integer value for
$s$, where $s(s+1)\equiv RA^{-2}$. The
solution, regular at $\tau\to-\infty$, is given by \cite{refl}
\bq
\psi(\tau)=C (1-\xi^2)^{\epsilon\over 2}
F\Bigl(\epsilon-s,\epsilon+s+1;\epsilon+1;\frac{1}{2}(1+\xi)\Bigr)
\label{solu}
\eq
where $C$ is determined by the initial conditions on $f_2(x)$,
$\epsilon=-E/A$, $\xi=tanh(A\tau)$ and $F(a,b;c;z)$ is the
hypergeometric function.
The number of the poles or bound states, leading to
non-zero amplitudes, is exactly equal to $s$ and the location of the
poles is given by
$E(k)=Ak, k=1,...,s$. This means that the only non-zero
amplitudes occur when
\bq
n=2Ak\;,\;\; k=1,\ldots,s\;,\;\;\;{\cal A}(2\to n)\neq 0\;\;.
\label{pole}
\eq
\par
It is possible to eliminate the $\tau$ dependence from
Eqs.(\ref{refp}) and (\ref{tau})
and obtain an equation wich involves only the potential
and its derivatives:
\bq
\biggl({U'''(\phi)\over 2A(1-U''(\phi))}\biggr)^2={1-(1-U''(\phi))R^{-1}
\over 2U}.
\label{poteq}
\eq
where $U(\phi)\equiv\frac{1}{2}\phi^2+V(\phi)$.
It is easy to see that, if for a potential $U(\phi)$, \eqn{poteq}
is satisfied with $R$ and $A$ such that $R/A^2=s(s+1)$, $s$ being
an integer, then the \a2n amplitudes at threshold vanish for any
$n$ except for those given by \eqn{pole}. The inverse is also true,
in the following sense: if \a2n amplitudes satisfy \eqn{pole},
then a potential $U(\phi)$ can be constructed, using Eqs.(\ref{refp})
and (\ref{tau}), which satisfies \eqn{poteq}.

\par The above result enables us to find
potentials with the nullification property. As a first,
rather trivial, application, we examine the case of a monomial
interaction, $U(\phi)=\frac{1}{2} \phi^2 + \frac{1}{m!}\phi^m$.
We find $R=m(m-1)/2$ and $A=(m-2)/2$, which gives
$s={m\over m-2}$. This is an integer only for $m=3,4$. For $m=3$,
$s=3$ and the amplitudes \a2n are non-zero only for $n=1,2,3$.
For $m=4$, $s=2$ and the amplitudes vanish except for $n=2,4$.
The case with $\phi^3$ as well as $\phi^4$
interactions is less trivial. The potential can
be written generally as $U(\phi)=\frac{1}{2} \phi^2+{\mu\over 6}
\phi^3 +\frac{1}{24} \phi^4$. The application of \eqn{poteq}
results in the determination of the constant $\mu$:
$\mu=\pm \sqrt{3}$. This corresponds to the broken-symmetry case
\cite{akp1}.
The latter is the only combination of $\phi^3$ and $\phi^4$ which
leads to zeros of \a2n amplitudes. In this case
$R=\frac{3}{2}$ and
$A=\frac{1}{2}$, so $s=2$ and the non-zero amplitudes \a2n
occur for $n=1,2$.
\par Equation (\ref{poteq}) enables us to go beyond these known examples.
We find that the sine-Gordon potential,
\bq
U(\phi)=(1-cos\phi)
\eq
satisfies \eqn{poteq} with
$A=1$ and $s=1$, so the only non-zero amplitude
is ${\cal A}(2\to2)$. This is a well known property in $d=2$
space-time dimensions, where the theory is integrable and the
nullification of \a2n, for $n\ge 4$,
holds for any kinematical configuration \cite{zamo},
due to the existence of an infinite set of conservation laws.
We find that the nullification survives in all space-time dimensions
at the kinemtaical threshold.
Furthermore the same nullification occurs for the sinh-Gordon
potential,
\bq
U(\phi)=(cosh\phi-1).
\eq
This can be verified directly, by solving the equation for $f(x)$
\cite{akp1}, which in the above cases is $f(x)=4tan^{-1}(\frac{x}{4})$
($=4tanh^{-1}(\frac{x}{4})$) and then verify that \eqn{difeq} is of the
form given by \eqn{ceneq}.
It is also possible to construct `customized' potentials in which
\a2n is nonzero at threshold for only one value of $n$, and vanishes
for all other values, by choosing appropriate values for $A$ and $s$
\cite{akpcoming}. Obviously, these potentials satisfy Eq.(\ref{poteq}).
\par
In the case where we have in the Lagrangian the $\phi$ field
coupled to other boson or fermion fields,
the nullification of ${\cal A}(\chi\chi\to n\phi)$
or ${\cal A}(f\bar f\to n\phi)$ has exactly the same explanation
in terms of reflectionless potentials \cite{vol4}. The only
difference is that in these cases the nullification
occurs when the couplings and/or the masses of the particles
obey definite relations among them. For instance, taking the coupling
${\cal L}_{int}=\frac{1}{4}\chi^2\phi^2+\frac{1}{24}\lambda_4\phi^4$
we find that for $g/\lambda_4=\frac{1}{6}s(s+1)$,
${\cal A}(\chi\chi\to n\phi)$ vanish for $n\ge 2(s+1)$.
This suggests that the search for
a generalization
of \eqn{poteq} in the case of several fields is undoubtedly interesting.
\par
The nullification we study here does not exhaust all the possible
cases where  \a2n
amplitudes vanish at threshold. For instance
the unitarity-respecting toy-model potential described in
\cite{akp3} does not satisfy \eqn{poteq}. Indeed,
although we have found vanishing \a2n amplitudes for all
$n$ even and larger than 4, there is no nullification of the
type discussed here, since the amplitudes are non-vanishing for
all odd values of $n$.
\par
It is worthwhile to note that \eqn{ceneq} is the same as
the equation describing the stability of 1+1 dimensional solitons
and is connected to scattering in the presence of solitonic backgrounds.
It seems therefore that the nullification of ${\cal A}(2\to n)$
amplitudes at threshold is a genuine dynamical effect, suggesting
that for certain theories, some integrability properties emerge
in the threshold kinematical configuration.
The study of the (as yet ill-understood) relation between nullification
at kinematical threshold and integrability in $d=2$ space-time
dimensions could provide us with a new insight into multiboson
production processes.

\end{document}